\documentclass[11pt]{article}
\pdfoutput=1

\usepackage[margin=1in]{geometry}
\usepackage[utf8]{inputenc}
\usepackage{latexsym}
\usepackage[T1]{fontenc}
\usepackage[utf8]{inputenc}
\usepackage{enumerate}
\usepackage[english]{babel}
\usepackage[utf8]{inputenc}
\usepackage{tcolorbox}
\usepackage{amsmath}
\usepackage{graphicx}
\usepackage[colorinlistoftodos]{todonotes}
\usepackage{indentfirst}
\usepackage{commath}
\usepackage{float}
\usepackage{multirow}
\usepackage{longtable}
\usepackage{multicol}
\usepackage{caption}
\usepackage{subcaption}
\usepackage{pdfpages}
\usepackage[utf8]{inputenc}
\usepackage[english]{babel}
\usepackage{fancyhdr}
\usepackage[hyphens]{url}
\usepackage{ragged2e}
\usepackage{amsmath}
\usepackage{adjustbox}
\usepackage{array}
\usepackage{caption}
\usepackage{subcaption}
\usepackage{lipsum}
\usepackage{dirtytalk}
\usepackage{wrapfig}
\usepackage{color}
\usepackage{bm}
\usepackage{authblk}

\title{A Computational Topology-based Spatiotemporal Analysis
Technique for Honeybee Aggregation}

\author[1]{Golnar Gharooni-Fard*}
\author[1]{Morgan Byers*}
\author[1]{Varad Deshmukh*}
\author[1,2]{Elizabeth Bradley}
\author[3]{Carissa Mayo}
\author[4,5]{Chad Topaz}
\author[1,2,6]{Orit Peleg}

\affil[1]{Department of Computer Science, University of Colorado Boulder, CO, USA}
\affil[2]{Santa Fe Institute, NM, USA}
\affil[3]{Department of Mathematics, University of Colorado, Boulder, CO, USA}
\affil[4]{Department of Applied Mathematics, University of Colorado, Boulder, CO, USA}
\affil[5]{Williams College, Williamstown, MA, USA}
\affil[6]{BioFrontiers Institute, University of Colorado Boulder, CO, USA}

\fancypagestyle{firstpage}
{
    \fancyhead[L]{Under review in \textit{npj Complexity}.}    
}

\date{}
\begin{document}
\maketitle
\thispagestyle{firstpage}

\begin{quote}
*Contributed equally
\end{quote}

\begin{quote}
{\noindent \bf Abstract:} A primary challenge in understanding
collective behavior is characterizing the spatiotemporal dynamics of
the group.  We employ topological data analysis to explore the
structure of honeybee aggregations that form during trophallaxis,
which is the direct exchange of food among nestmates.  From the
positions of individual bees, we build topological summaries called
CROCKER matrices to track the morphology of the group as a function of
scale and time.
Each column of a CROCKER matrix records the number of topological
features, such as the number of components or holes, that exist in the
data for a range of analysis scales at a given point in time.  To
detect important changes in the morphology of the group from this
information, we first apply dimensionality reduction techniques to
these matrices and then use classic clustering and change-point
detection algorithms on the resulting scalar data.  A test of this
methodology on synthetic data from an agent-based model of honeybees
and their trophallaxis behavior shows two distinct phases: a dispersed
phase that occurs before food is introduced, followed by a
food-exchange phase during which aggregations form.  We then move to
laboratory data, successfully detecting the same two phases across
multiple experiments.  Interestingly, our method reveals an additional
phase change towards the end of the experiments, suggesting the
possibility of another dispersed phase that follows the food-exchange
phase.

\end{quote}

\section{Introduction}
\label{sec:intro}

\begin{figure}
\begin{center}
\includegraphics[width=0.6\textwidth]{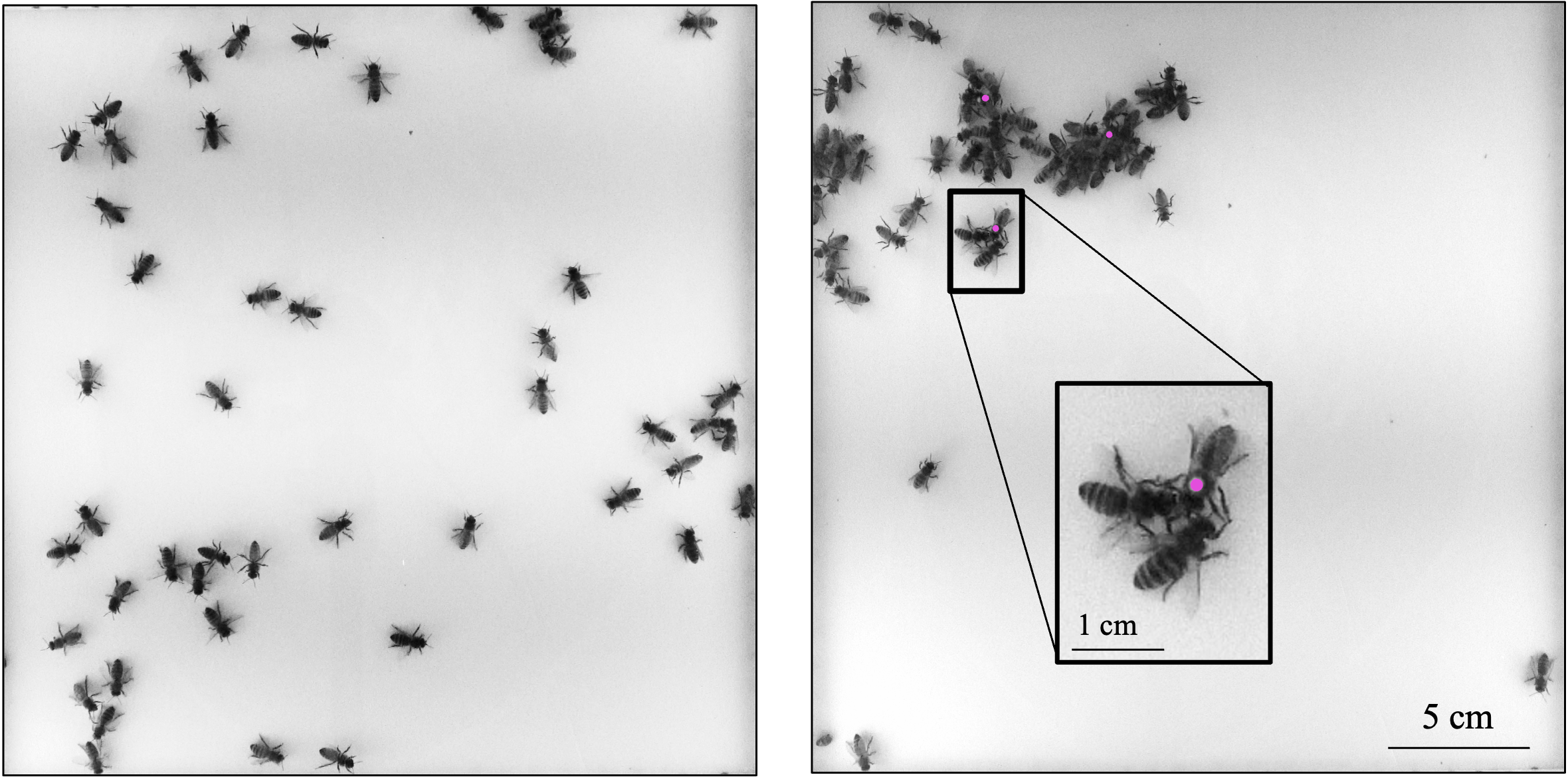}
\\
(A) \hspace*{1.5truein} (B)
\end{center}
\caption{Trophallaxis in honeybees.  In this experiment, a number of
  fed {\sl donor} bees are introduced into a larger group of {\sl
    food-deprived} bees, at which point they begin exchanging food and
  form aggregations.  Panel A shows the situation at the beginning of
  the experiment ($t=0$), before the introduction of the donor bees at
  $t=430$ seconds.  Panel B shows aggregations that have formed by
  $t=545$, with an inset focusing on a donor bee (marked with a pink
  dot) and two receiver bees as they exchange food.  }
\label{fig:setup}
\end{figure}

The sophisticated social organization among honeybees ({\sl Apis
  mellifera L.}) involves intricate interaction networks that are
critical to the function of the hive.  An important instance of this
is the exchange of food
among colony members, which is performed through a process called
trophallaxis \cite{leboeuf2017trophallaxis}, a mutual feeding technique that
involves direct transfer of liquid food among nestmates.  Trophallaxis
interactions cause aggregations to form in the group
\cite{fard2020data}, as shown in Fig.~\ref{fig:setup}.  Analysis of
these patterns can help us understand how this collective
food-exchange network evolves over time, which can in turn provide
valuable information about the efficiency of global food distribution
among honeybees.

In this paper, we use topological data analysis (TDA) to perform a
rigorous spatiotemporal analysis of the morphology of honeybee groups
during the process of trophallaxis.  TDA is a set of mathematical
tools for characterizing the shape of real-world data.  One of those
tools, known as {\sl persistent homology}
\cite{Edelsbrunner02,Robins02}, takes a variable-resolution approach
to the shape-analysis problem, treating points as connected if they
are within some distance $\epsilon$ of one another and counting the
number of topological features---connected components, two-dimensional
holes, three-dimensional voids, etc.---in the resulting {\sl
  simplicial complex}.  By repeating that analysis for a range of
values of the $\epsilon$ parameter, this method produces a rich,
multi-scale signature of the spatial structure of a point cloud.  This
approach has grown in popularity for a variety of applications over
the past decade, including a number of biology problems
\cite{mcguirl2020topological,amezquita2020shape,topaz2015topological,
  loughrey2021topology,ciocanel2021topological,ulmer2019topological,bhaskar2019analyzing},
but it has not yet been applied to honeybee groups, whose patterns
have largely been studied using computer-vision techniques
\cite{fard2020data,nguyen2021flow,szopek2013dynamics}.  These require
     {\sl a priori} definitions of the specific scales at which
     aggregation occurs, however.  TDA offers a reliable way to
     perform a detailed spatiotemporal analysis of the aggregations
     without any such assumptions.

Here, we use the tools of TDA to study honeybee aggregations in the
context of trophallaxis, with a particular emphasis on discerning how
these aggregations evolve over time.  Our approach draws an analogy
between changes in these patterns, termed {\bf phase changes} in the
rest of this document, and the density phase transitions observed in
condensed matter \cite{ong2001,anderson2002,Shibayama1993}.  Notable
density-related phases in this application encompass a sparse phase
(where bees are uniformly distributed across the arena), a dense phase
(where bees form a cohesive cluster), as well as various intermediate
states characterized by combinations of dense and sparse clusters,
such as the presence of multiple smaller clusters or the coexistence
of dense and sparse configurations.  In order to track the dynamical
evolution of the topological signatures produced during our analysis
of these changing patterns, we use the CROCKER method (Contour
Realization Of Computed k-dimensional hole Evolution in the Rips
complex), a matrix-based representation that captures the morphology
of a point cloud as a function of both scale and
time~\cite{topaz2015topological}.  Each column of a CROCKER matrix
corresponds to a particular time point in the experiment.  The
elements of that column vector record the number of topological
features that exist in that data snapshot for a range of analysis
scales---{\sl e.g.}, the number of $\epsilon$-connected components for
a range of values of the scale parameter $\epsilon$.  We use
dimensional reduction techniques to convert each of these vectors to a
scalar, then use clustering algorithms to find the phase changes in
the resulting time series.

In the following section, we describe our methods in more detail and
demonstrate them using a synthetic dataset from an agent-based model
of trophallaxis in honeybees.  In Section~\ref{sec:experiments}, we
use this methodology to study trophallaxis in a laboratory experiment.
In both simulated and real data, these methods clearly bring out the
different phases in the behavior.  We discuss some alternative
approaches and implications of our findings in
Section~\ref{sec:discussion}.  Throughout this document, we use the
term ``aggregation'' to describe a group of bees in physical space and
the term ``clustering'' for the action of algorithms that find groups
in the resulting structure.

\section{Spatiotemporal TDA of honeybee aggregations}
\label{sec:tda}

\begin{figure}
\begin{center}
\includegraphics[width=0.8\textwidth]{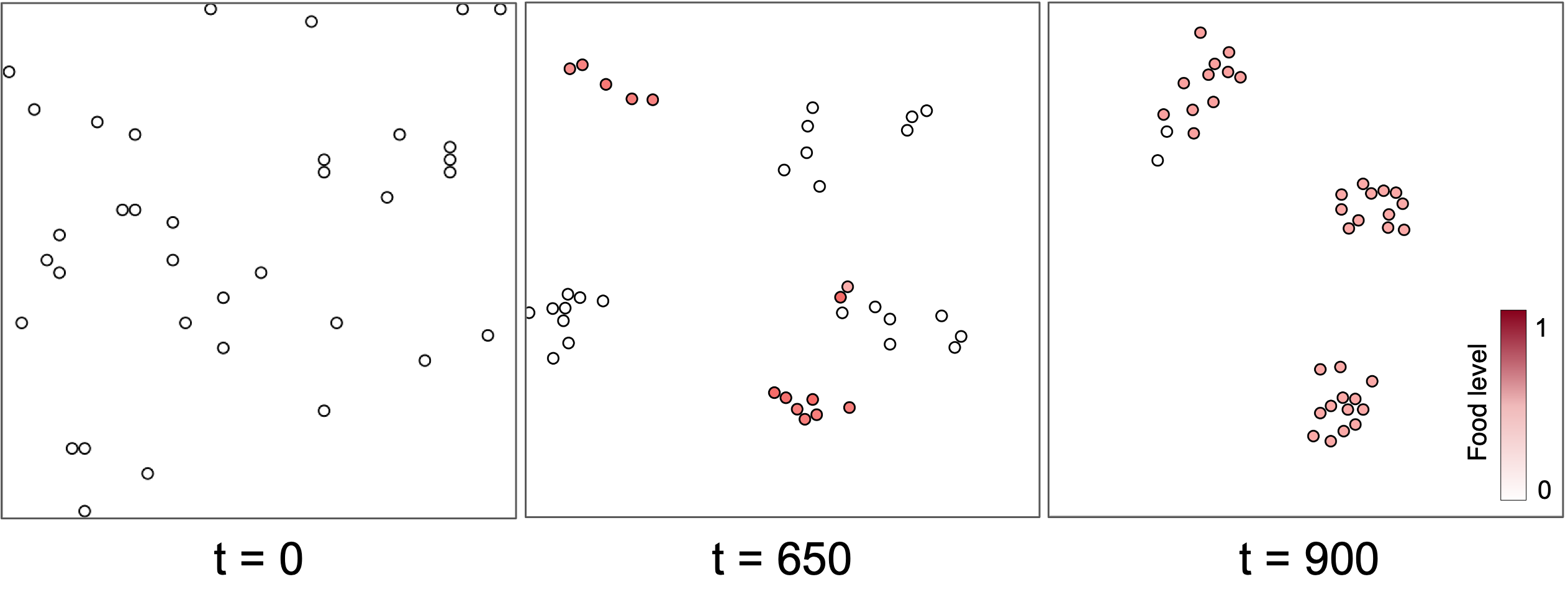}
\\
(A) \hspace*{1.3truein} (B) \hspace*{1.3truein} (C)
\end{center}
\caption{Three snapshots from an agent-based model of trophallaxis: at
  the start of the simulation ($t=0$), a 36 by 36 cell arena with
  reflective boundary conditions contains 38 food-deprived agents
  moving via random walk.  Four donor bee-agents are introduced at
  $t=400$, after which aggregations form as the food carried by those
  agents is distributed across the group.  Agents are colored by their
  amount of food (zero food shown in white and maximal food capacity
  shown in red).  Please see the Supplementary Materials for a video
  from a representative simulation run.}
\label{fig:abm-model}
\end{figure}

As a demonstration case for this discussion, we use an agent-based
model of trophallaxis that was first described in~\cite{fard2020data}.
To avoid repetition, we provide a concise summary of the simulation;
please see \cite{fard2020data} for more details.  At the start of each
run of this model, a number of bee-agents with zero values of a
food-level variable are placed randomly in a 2D simulation arena, as
depicted in Fig.~\ref{fig:abm-model}(A).
As the simulation progresses, these agents perform random walks,
remaining scattered around the arena.  Partway through the simulation
run, a number of donor bee-agents---with positive values of the
food-level variable--are introduced into the arena.
All agents continue moving via random walks, stopping to exchange food
with one another if they come within a predefined attraction radius
(2.5 simulation patches).  The length of the food exchange, during
which both agents remain stationary, is proportional to the difference
between their food levels; at its end, the food levels are equalized
between the two.  During this process, aggregations form in the group,
as shown in Fig.~\ref{fig:abm-model}(B) and~(C), while the food is
distributed across the agents.  For this series of experiments, we end
all model runs at $t=900$.

\begin{figure*}[t]
\begin{center}
\includegraphics[width=\textwidth]{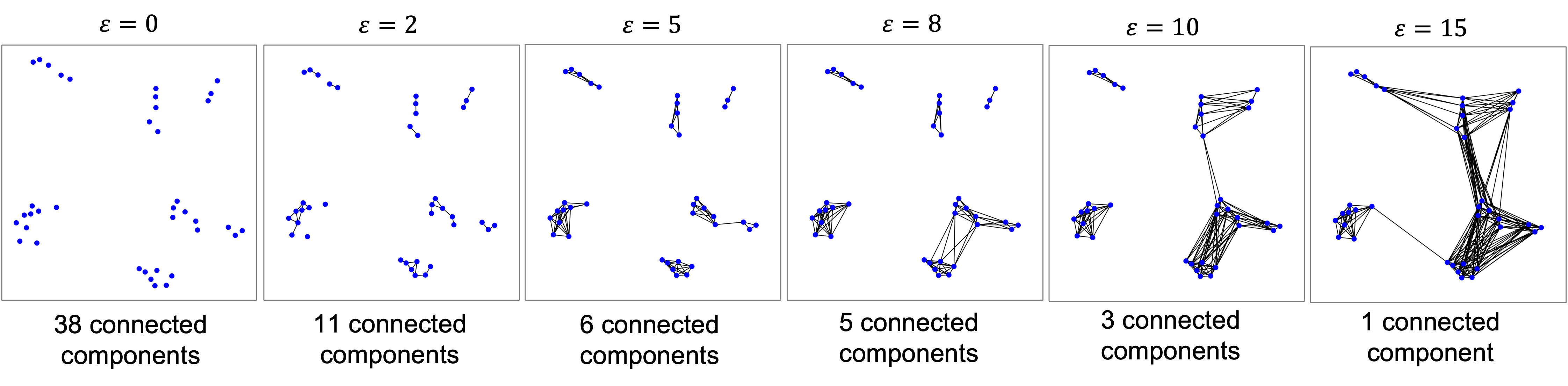}
\end{center}
\caption{A series of Vietoris-Rips complexes constructed from the
  positions of the bees in Fig.~\ref{fig:abm-model}(B) for six values
  of the filtration parameter $\epsilon$. 
}
\label{fig:rips}
\end{figure*}

Note that the morphology of the group of agents cannot be simply
classified as ``aggregated'' or ``not-aggregated.''  The structure in
the three panels shows different degrees of aggregation; moreover, any
classification as to the degree of aggregation will depend on what
measure of proximity one uses to define membership in an aggregation.
Topological data analysis is an effective way to quantify the
multi-scale nature of this structure in an effective and formal way.
As mentioned previously, persistent homology characterizes the shape
of a data set by analyzing it at different resolutions.  This involves
building a {\sl filtration} from the point cloud: a series of
simplicial complexes that capture its structure for different values
of a resolution parameter, $\epsilon$.  For this purpose, we use the
Vietoris-Rips approach \cite{vietoris-rips}, which constructs a
complex by creating balls of radius $\epsilon$ around each data point.
A set of $m$ points is connected by an $m$-simplex if every pair of
points in the set have intersecting $\epsilon$-balls.
From each of the resulting series of simplicial complexes, one then
computes the {\sl Betti numbers}: $\beta = \lbrace \beta_0, \beta_1,
\beta_2, .... \rbrace$, where $\beta_0$ is the number of connected
components, $\beta_1$ is the number of two-dimensional loops,
$\beta_2$ is the number of three-dimensional voids, and so on.
Fig.~\ref{fig:rips} shows an example filtration: a series of
simplicial complexes constructed from the data in
Fig.~\ref{fig:abm-model}(A) for six values of $\epsilon$.  The effects
of the filtration parameter are clearly visible across the panels of
the figure: for $\epsilon$ = 0, each agent is its own connected
component---{\sl i.e.}, $\beta_0=38$, the number of agents in the
simulation at that point in time---while for $\epsilon=15$ all agents
are connected together in one connected component ($\beta_0=1$).  In
between those values, the component structure reflects the patterns in
the spacing of the bees as the $\epsilon$ value grows to span larger
and larger gaps between the individuals, connecting them in the
Vietoris-Rips complex.  Values for the other Betti numbers $\beta_k$
can be similarly calculated for different $\epsilon$ values to produce
a multi-scale topological signature of the point cloud.

To carry out these calculations, one must specify the scales for the
construction: specifically, the range $[\epsilon_{min}, \,
  \epsilon_{max}]$ and spacing $\Delta \epsilon$ of the filtration
parameter.  A common approach to choosing the range is to take
$\epsilon_{min}$ at the value where each point is its own component
and $\epsilon_{max}$ such that the entire set is connected.  For the
example in Fig.~\ref{fig:rips}, this approach suggests
$[\epsilon_{min}, \, \epsilon_{max}]=[0, \, 15]$.  In other runs of
the model, however, a higher $\epsilon$ was required to connect all
the agents into a single component, so we standardize by using
$\epsilon_{max}=20$, which was adequate to produce
$\beta_0(\epsilon_{max})=1$ for all time points in all model runs.
Choosing the spacing, $\Delta \epsilon$, involves balancing
computational complexity, analysis resolution, and the scales in the
data.  What one wants is a $\Delta \epsilon$ that yields new
information at each step: if it is too small, the filtration will
contain multiple elements with identical topology; if it is too large,
those elements may skip over $\epsilon$ values where important
topological changes occur.  To steer between these extremes, we
calculate the average pairwise distances between bees in neighboring
cells across every time point in every model run, obtaining a value of
1.41, and then take a somewhat smaller value $\Delta \epsilon =1$ to
be sure not to miss important topological changes.  This choice yields
$n=21$ simplicial complexes at each time point.  Since we are
interested in aggregations, we focus on the number of
$\epsilon$-connected components, $\beta_0$, in each of these
complexes.  This computation, which we perform with the {\tt GUDHI}
Python package~\cite{gudhi:urm, gudhi:RipsComplex}, requires 62
$\mu$sec for each of the 21 complexes in the filtration ({\sl i.e.},
0.0013 sec total for each time point) on an Apple M1 Pro with 10 CPU
cores and 16GB of main memory.

\begin{figure}
\begin{center}
\includegraphics[width=0.6\linewidth]{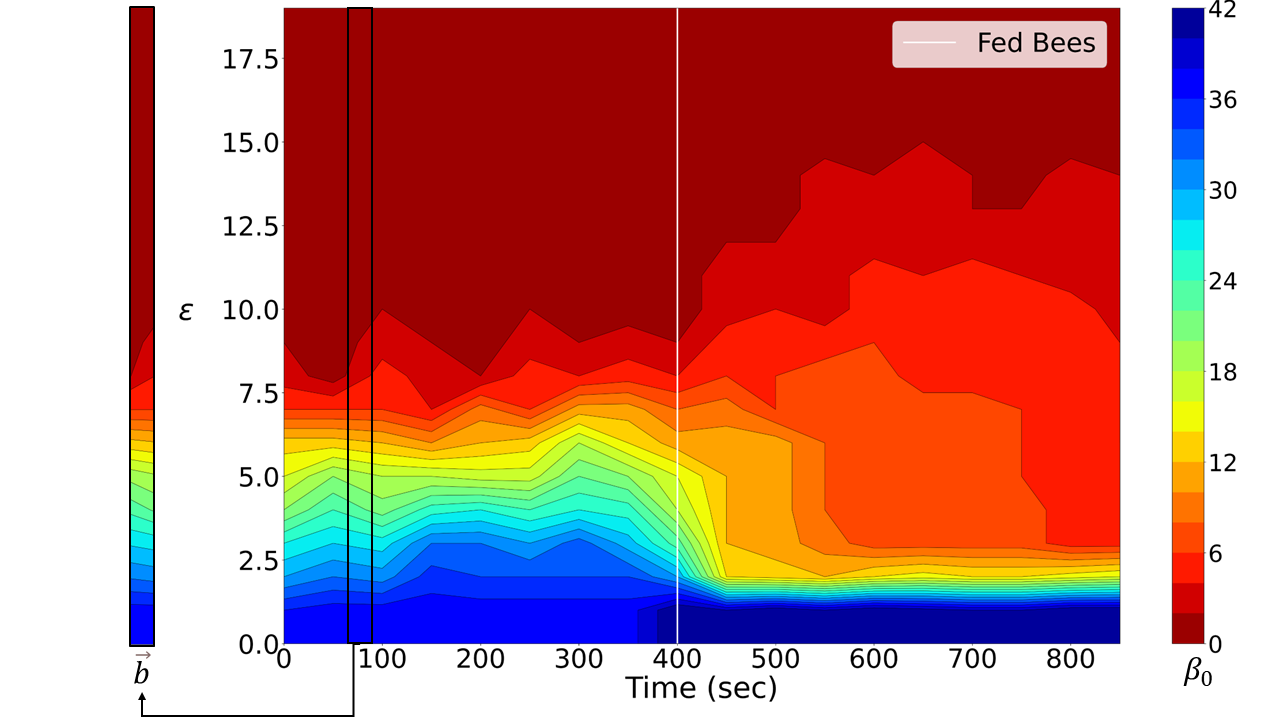}
\\
\medskip
\hspace*{7mm}
\includegraphics[width=0.6\linewidth]{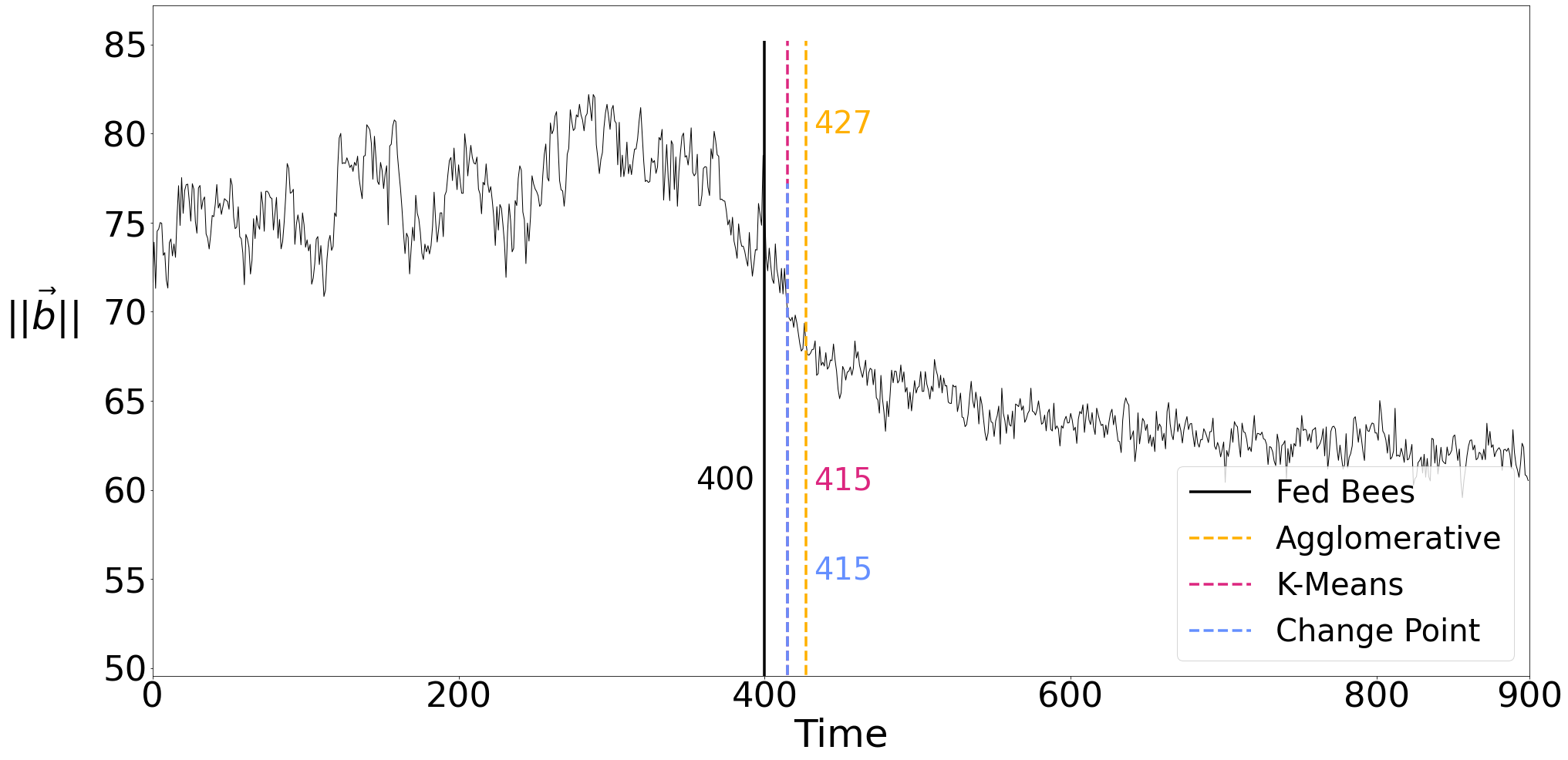}
\end{center}
\caption{Top: a CROCKER plot of the positions of the bees in an
  agent-based model run.  The colors indicate the number of connected
  components ($\beta_0$) in the simplicial complex constructed with
  the $\epsilon$ value on the $y$ axis.  The black box highlights a
  vertical slice of the CROCKER plot, $\vec{b}= [ \beta_0(\epsilon_i)
  ] $, at the corresponding time point.  Bottom: a time series of the
  $\ell^2$ norms of each of these $\vec{b}$ vectors, with superimposed
  dashed lines for the results of the change-point detection
  algorithm, shown in red, and the two clustering algorithms:
  $k$-means in blue and agglomerative in orange.}
\label{fig:model_crocker}
\end{figure}

From a data-analysis standpoint, the procedure described above can be
viewed as converting a set of points into a $n$-vector
$[\beta_0(\epsilon_0), \beta_0(\epsilon_1), ... \beta_0(\epsilon_n)]$
that characterizes the shape of that point cloud for $n$ different
values $[\epsilon_1, \epsilon_2, ..., \, \epsilon_n]$ of the
resolution parameter.  To detect phase changes in that structure,
we need a way to track the evolution of that shape with time.  Various
representations have been proposed for that purpose, including CROCKER
plots~\cite{topaz2015topological}, vineyards~\cite{cohen2006vines},
multiparameter rank functions~\cite{Kim2021} and CROCKER
stacks~\cite{Xian2022}.  In this paper, we use CROCKER plots, which
are two-dimensional representations with time on the horizontal axis
and information about scale and structure on the other.  The columns
of the matrices that are rendered by these plots are a series of
vectors like the ones mentioned above, each of which records, for a
given point in time $t_i$, the value of $\beta_0$ for each of $n$
values of $\epsilon$ in the filtration: \[\vec{b}(t_i)= [\beta_0(t_i;
  \epsilon_0), \beta_0(t_i; \epsilon_1), ... \beta_0(t_i;
  \epsilon_n)]^T\]
\noindent The plots themselves visualize this information using
color-coded contours on the $\vec{b}$ values.  The top panel of
Fig.~\ref{fig:model_crocker} shows an example: a CROCKER plot
constructed from the bee positions in one run of the agent-based
model.
The dark blue region across the bottom of the image reflects the large
number of $\epsilon$-connected components that exist in the simplicial
complex when $\epsilon$ is small; the contour that divides the dark
brown region from the burnt-orange region identifies the $\epsilon$
value at which the complex at the associated time point includes all
of points in the simulation.  The plot reveals a clear shift in the
topology soon after the introduction of the donor bees, where the
contours change drastically around $t = 400 sec$.  Lower-valued
countours (separating red sections) separate and raise slightly, and
higher valued course fall and bunch together more closely.  Overall,
these changes mean that reduction from a large number of clusters to a
moderate number, say from 42 to around 10, occurs at smaller scales
than before.  Conversely, reduction from moderate numbers of clusters
to a single cluster occur at larger scales than before. These
observations are consistent with bees forming small groups spaced
throughout the domain.

This shift in the patterns confirms previous results cited in
Section~\ref{sec:intro} about the ability of CROCKER plots to make
phase changes in biological aggregations visually apparent.  The
goal of this paper is to go beyond visual observations and develop
formal methods for detecting those phase changes automatically.
We evaluate two different types of algorithms for the associated
calculations:
\begin{itemize}
  \item A standard change-point detection algorithm: recursive binary
    segmentation of the time series, based on a likelihood ratio
    test.
  \item Two unsupervised clustering algorithms---$k$-means and
    agglomerative~\cite{zbMATH03340881,Zepeda-Mendoza2013}---which are
    representative of this class of methods.
    \end{itemize}
The basic idea is to apply these algorithms to the column vectors
$\vec{b}(t)$ in the CROCKER matrix.  If the vectors before and after
the introduction of the donor bees do indeed capture some distinct
morphology, the algorithms should separate them into two separate
regimes, effectively resulting in a phase-change detection.  The column
vectors require some pre-processing before these algorithms can be
applied, however.  Their high dimensionality, coupled with the
comparatively short number of time points in each phase, can make it
difficult to run clustering algorithms---the change-point detection
algorithm requires a scalar input.  To work around this, we apply
dimensional reduction techniques to the column vectors and then run
the phase-shift detection algorithms on the resulting low-dimensional
time series.  A simple way to do this is to take the $\ell^2$ norms of
each $\vec{b}(t)$.  The results of this procedure, applied to the
CROCKER matrix plotted in the top panel of
Fig.~\ref{fig:model_crocker}, are shown in the bottom panel of that
figure.  There is a clear shift in the normed values at the time of
the change in structure of the CROCKER plot after the introdution of
the donor bees.

To isolate this shift, we apply the three different algorithms to this
scalar time series, with the goal of identifying the time $t_{shift}$
at which the structure changes.  In the $k$-means algorithm, the user
must specify how many clusters ($k$) to search for.  Since we are
looking for two distinct phases in the $\vec{b}$ vectors, we choose
$k=2$.  The algorithm begins by randomly mapping each vector in the
dataset to one of the $k$ clusters, then computes the centroids of
those clusters.  The distance from these centroids to each vector is
computed and the vectors are re-assigned to the cluster with the
closest centroid.  This process is repeated until the algorithm
converges: {\sl i.e.}, when no vectors change cluster membership.
Agglomerative clustering is a bottom-up, hierarchical clustering
method: each vector starts as its own cluster, and then vectors are
successively grouped based on some linkage criterion---{\sl e.g.},
Ward's method \cite{Ward1963HierarchicalGT}, in which clusters are
merged in a way that minimizes the overall variance within each
cluster.  (Note that in this analysis we are not working with the full
$\vec{b}$ vectors, but rather with their norms $|| \vec{b} ||$, so all
of the calculations described above actually involve scalar values,
not multi-element vectors.)  We use the R implementation (``{\tt
  changepoint}'') of the recursive\footnote{Since we were looking for
a single change point, there was no need for the recursive step here.}
binary segmentation/likelihood ratio test algorithm
\cite{changepoint2014}, the {\tt scikit-learn} implementations of both
clustering algorithms \cite{scikit-learn}, and the $\ell^2$ norm for all
distance computations.

For this model run, all three algorithms yield very similar results,
flagging the phase change soon after the introduction of the donor
bees at $t= 400$: agglomerative at $t_{shift}=427$ and both {\tt
  changepoint} and $k$-means at $t_{shift}=415$ (shown superimposed on
the norm trace in the lower panel of Fig.~\ref{fig:model_crocker}).
This similarity persists across different model runs: for 50
repetitions of the numerical experiment, the detected $t_{shift}$
values of the three algorithms were similar: 422.8, 429.0, and 422.2
for $k$-means, agglomerative, and {\tt changepoint}, respectively,
with standard deviations of 12.9, 27.2, and 12.9.
These lags are biologically sensible: since it takes some time for the
bees in the arena need to first sense the presence of the donors and
then cover the distance to them, there will always be a lag between
the introduction of the donors and the formation of aggregations.
Note that it is difficult to make any assertions about ground truth
here---aside from the obvious point that trophallaxis-induced
aggregation should not occur {\sl before} the introduction of the
donor bees---without performing more-detailed modeling of that
sense/move process.

Notably, both $k$-means and agglomerative algorithms yield clusters
that are well delineated in time: that is, nearly all norm values
$||\vec{b}(t)||$ that occur before the phase change $t_{shift}$
are assigned to the first cluster and nearly all $||\vec{b}(t)||$ that
occur after $t_{shift}$ are assigned to the second cluster.  The
delineation is not perfect; rather, there is generally a short span of
time where the classification of the $||\vec{b}(t)||$ values
alternated between the two clusters.  To formalize this, we define an
{\sl overlap region} $[t_{left},t_{right}]$, where $t_{left}$ is the
first data point that is classified as a member of the second cluster
and $t_{right}$ is the the last data point that is classified as a
member of the first cluster\footnote{As an example, consider the
series of norm values labeled as follows:
0000000000101100010111111111.  Here, the boundaries of the overlap
region fall at $0000000000 \, | \, 101100010 \, | \, 111111111$ and
its width is nine.}.  When overlap occurs, we take the cluster
delineation to fall at the first time step at which no more
``mislabeled'' vectors occur ({\sl i.e.}, $t_{shift}=t_{right}$).
Across the different model runs, the maximum width of the overlap
region was five time steps, with an average of 1.75.  In view of the
fact that time is not included explicitly in the calculations---recall
that only the norm values are passed as inputs to the clustering
algorithms---a delineation this crisp is an encouraging result, as it
suggests that the topological signature truly captures the salient
features of the structure.  In Section~\ref{sec:experiments}, we offer
another approach that annotates each normed value with the associated
time stamp---{\sl i.e.}, running the clustering algorithms on a set of
vectors $[\, ||\vec{b}(t_i) ||, t_i]$ rather than the scalars $||
\vec{b}(t_i) ||$---which proves to be useful when the data are not so
clean.  In Section~\ref{sec:discussion}, we explore the alternative of
principal component analysis for dimensional reduction.

\section{Applications to Experimental Data}
\label{sec:experiments}

Quantifying the morphology of synthetic data is a useful first test,
but the ultimate goal of our work is to understand trophallaxis in
real honeybees.  To that end, we apply our method to data collected
via video from a series of six experiments in a semi-2D arena that
measures 36 cm $\times$ 36 cm $\times$ 2 cm.  At the start of each
experiment, the arena contains a group of honeybees that had been
deprived of food for 24 hours.  After recording their behavior for
several minutes, we introduce multiple donor bees that had retained
free access to food until that point.  We then continue recording the
group motion for about 30 minutes, as the bees interact and exchange
food.  Video frames from these experiments are greyscale images 1400
pixels on a side at a resolution of 60 pixels per centimeter.  These
data are recorded at 30 frames per second, a rate that far exceed the
temporal resolution necessary to completely capture the motion of the
bees, so we downsample to a rate of one frame per second for the
following analysis.  Since the pixel size in these images dictates the
lower bound on the spatial resolution of the morphological analysis,
as well as the quantization of the range of that analysis, we use
pixels as the fundamental distance unit in the analysis that follows,
rather than mks units or body-length scales of the animals.  The
implications of this are discussed further in
Section~\ref{sec:discussion}.

\begin{figure}
\begin{center}
\includegraphics[width=\linewidth]{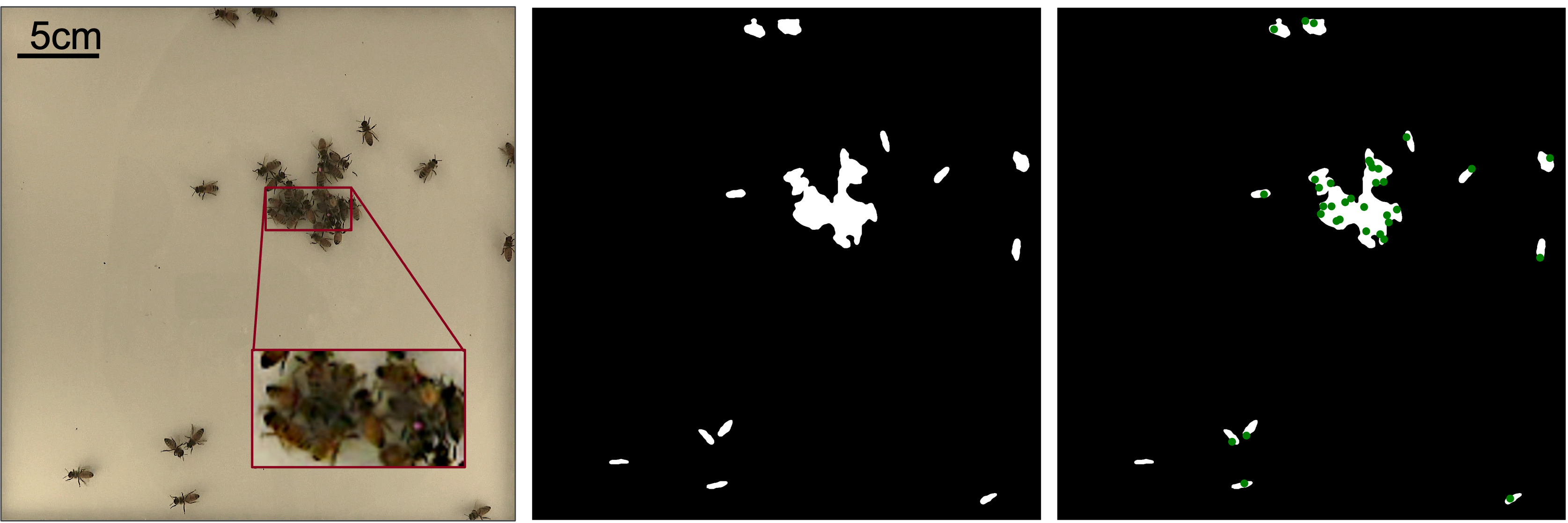}
\\
(A) \hspace*{2.3truein} (B) \hspace*{2.3truein} (C)
\end{center}
\caption{Building a point cloud of bee positions: (A) Video frame with
  inset showing detailed structure. (B) Segmented image containing all
  pixels occupied by bees (C) Point cloud.}
\label{fig:build-point-cloud}
\end{figure}

\begin{figure}
\begin{center}
\includegraphics[height=0.3\linewidth]{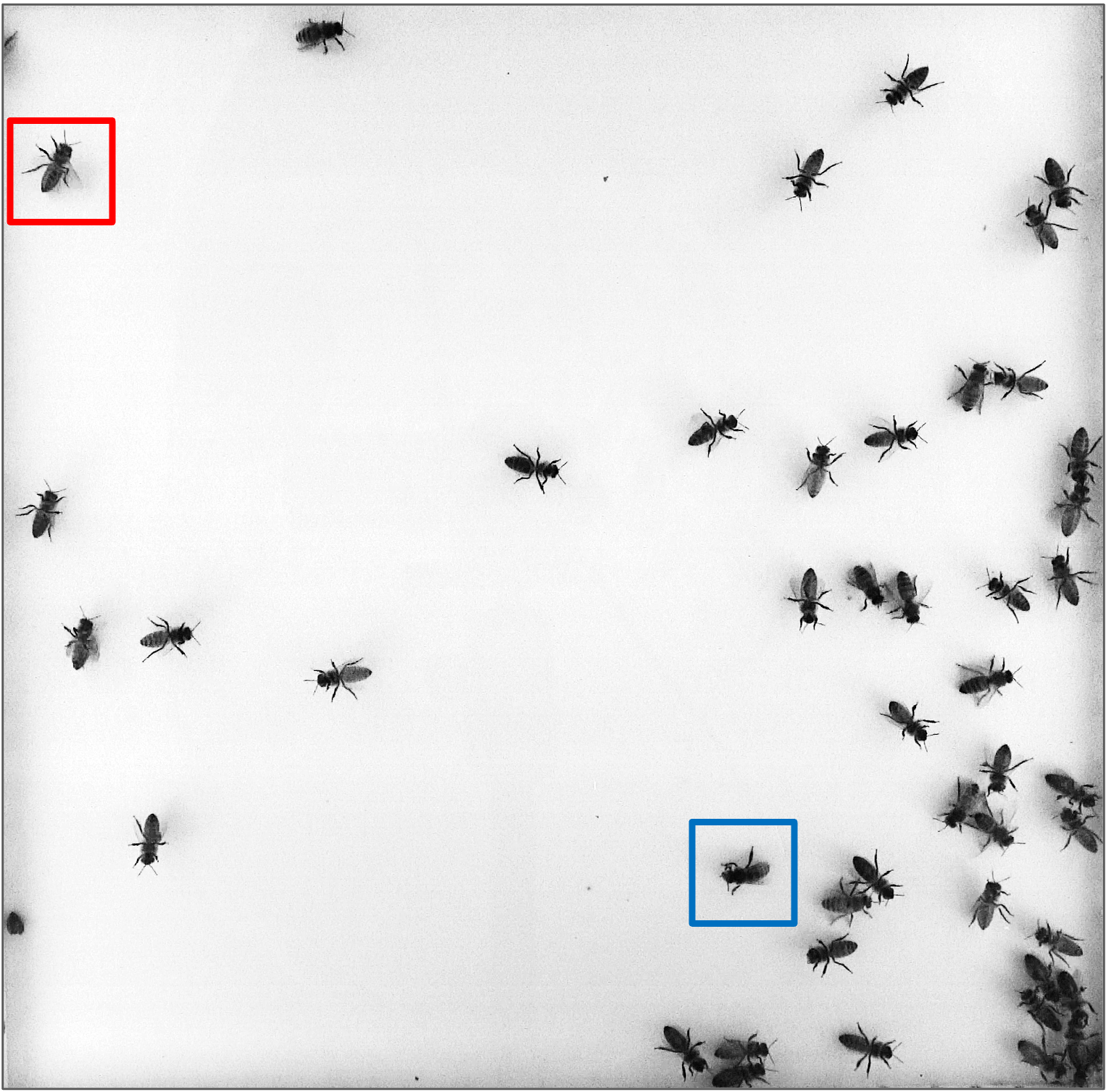}
\hspace*{3mm}
\includegraphics[height=0.3\linewidth]{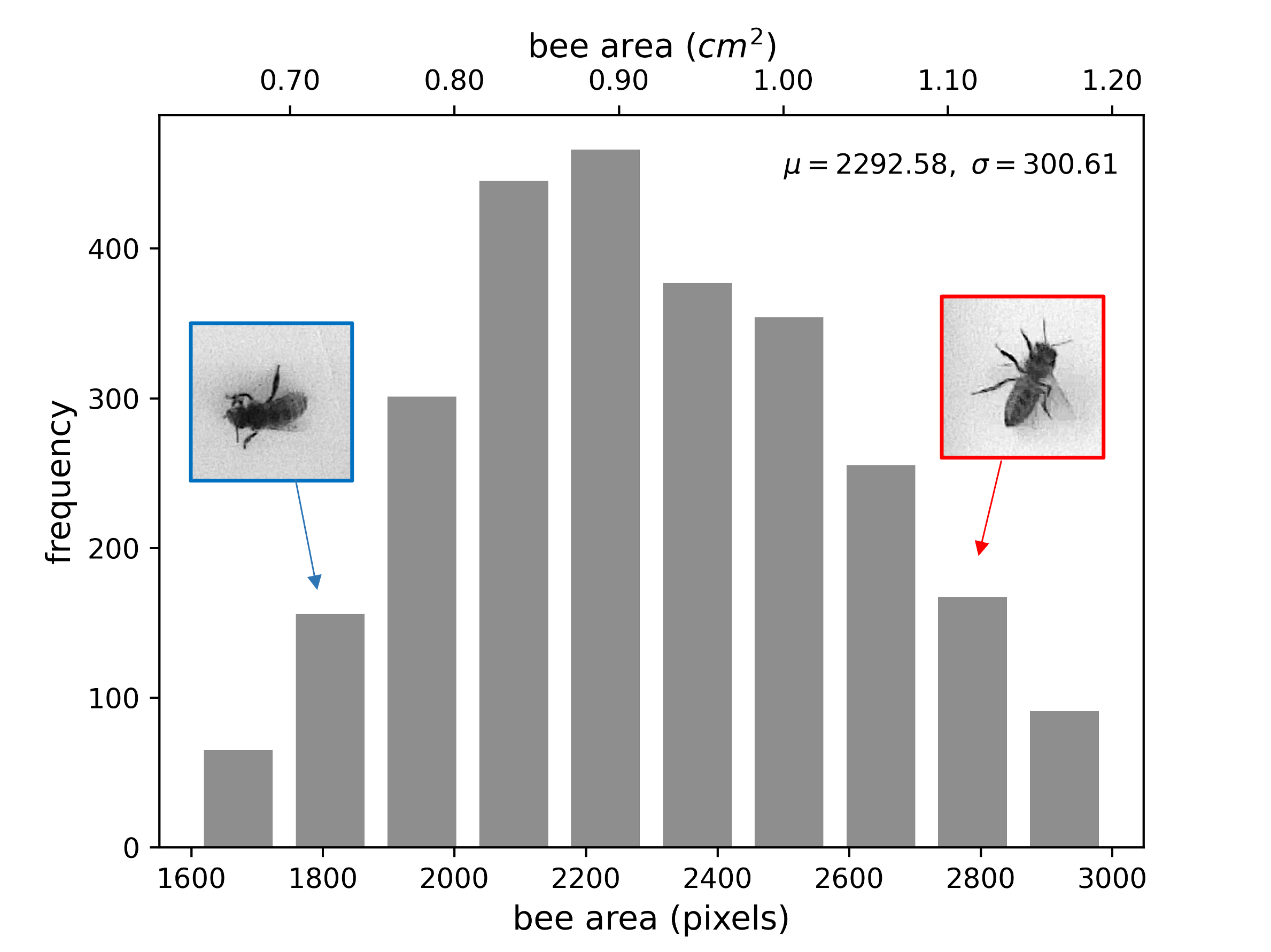} 
\\
(A) \hspace*{2.5truein} (B)
\end{center}
\caption{Body size of the honeybees in the experiments: (A) Example
  frame. (B) Distribution of bee sizes in 100 frames.}
\label{fig:bee-size}
\end{figure}

Before we can use TDA to analyze these data, we must first build a
point cloud from each frame of these videos.  Despite recent
improvements \cite{bozek2021markerless, ngo2019real}, this is
difficult because the bees touch and even occlude one another.  This
makes it hard for an algorithm to detect the exact positions of
individuals in images like these.  One can address this by affixing
barcodes to each bee, as in~\cite{gernat2018automated}, or by
hand-labeling each image, but those approaches are onerous and
expensive, especially when one has 30$+$ minutes of 30 fps video.
Instead, we image segmentation techniques to approximate the positions
of the individual bees.  To each image, we first apply a Gaussian blur
filter with a $7 \times 7$ pixel kernel to reduce background noise
\cite{nixon2019feature}, then employ local minimum/Otsu
thresholding~\cite{otsu1979threshold} to distinguish the darker pixels
({\sl i.e.}, bees) from the lighter background.  Since the resulting
images are still speckled with random false-positive pixels, we use
erosion and closure \cite{maragos1999morphological} to remove them.
This produces an image like the one shown in
Fig.~\ref{fig:build-point-cloud}(B), with pixels corresponding to bees
shown in white and background in black.  The next step is to aggregate
the contiguous white pixels in the image into groups, with adjacency
defined as a shared edge or a corner.  We then determine the number of
bees in each of these groups by dividing its area (in pixels), by the
average area in pixels of a single bee, which we obtain from a
separate series of experiments; see Fig.~\ref{fig:bee-size}.
Finally, we create the point cloud by distributing that number of
points randomly inside the boundaries of the group, as shown in
Fig.~\ref{fig:build-point-cloud}(C).  

\begin{figure}
\begin{center}
\includegraphics[width=0.6\linewidth]{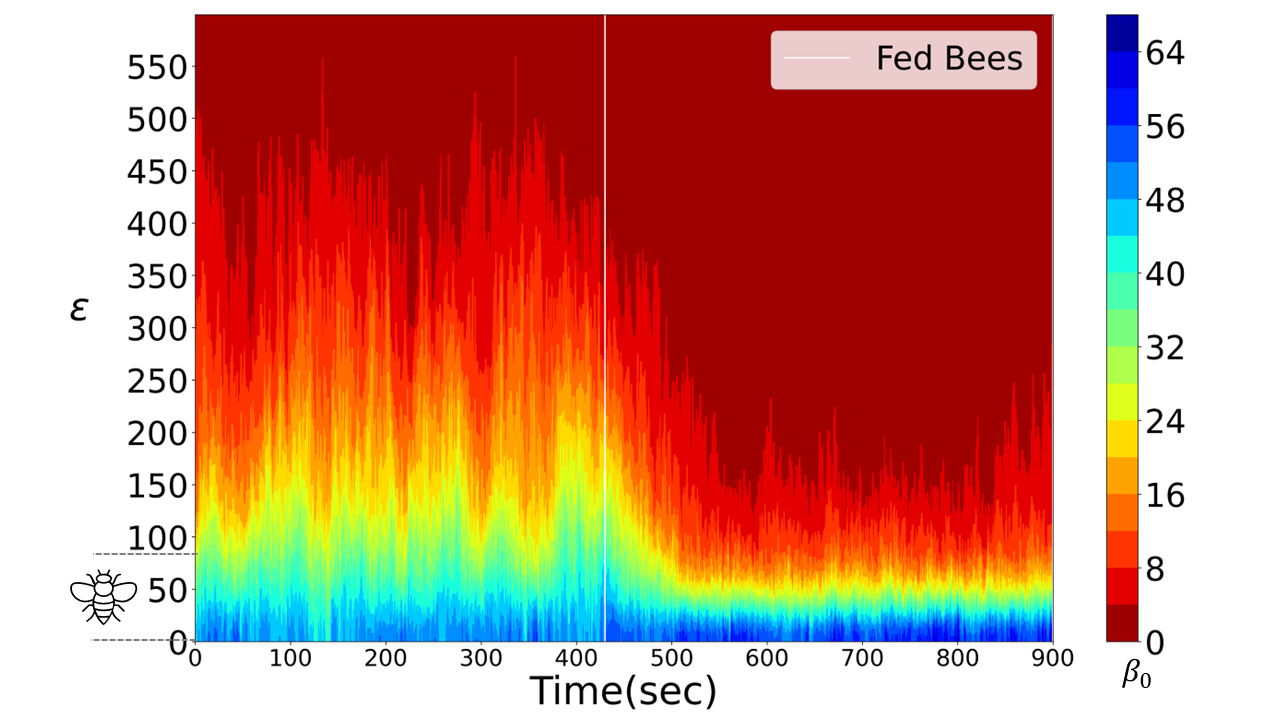}
\\
\medskip
\includegraphics[width=0.6\linewidth]{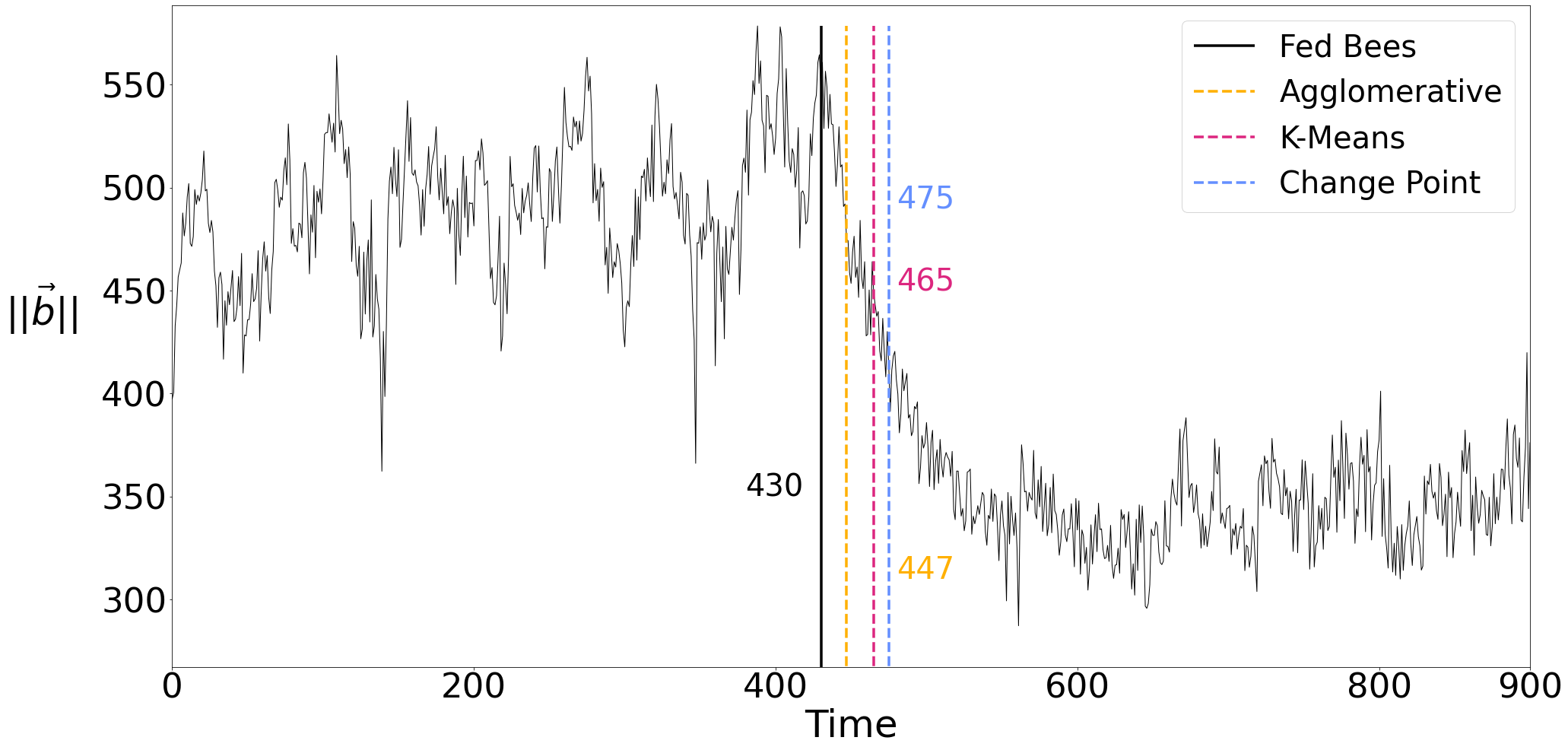}\hspace*{10mm}
\end{center}
\caption{Top: a CROCKER plot from a laboratory food-exchange
  experiment like the one pictured in Fig.~\ref{fig:setup}.  Computing
  each column in this plot requires 0.07 seconds on a Mac 1.7 GHz
  Quad-Core Intel Core i7 and 16 GB of memory---significantly longer
  than in Fig.~\ref{fig:model_crocker} because of the much larger
  number of $\epsilon$ values in the filtration.  Bottom: the
  corresponding $|| \vec{b}(t_i) ||$ time series, with superimposed
  dashed lines for the results of the change-point detection
  algorithm, shown in blue, and the two clustering
  algorithms---$k$-means (red) and agglomerative (orange)---applied to
  the time-annotated norm values $[\, ||\vec{b}(t_i)||,
    t_i]$.}
\label{fig:exp_crocker}
\end{figure}

The analysis of these point clouds proceeds as described in the
previous section, beginning with the construction of a series of
Vietoris-Rips complexes from the point cloud at each time $t_i$.  An
$\epsilon$ value equal to 700 pixels in the experimental video
frames---{\sl i.e.,} $\approx$ 12.7cm or eight bee body lengths---is
adequate to connect all points in the cloud into a single connected
component, so we use $[\epsilon_{min}, \, \epsilon_{max}]=[1, \, 700]$
with $\Delta \epsilon=1$, all in the units of pixels.
As in the analysis of Section~\ref{sec:tda}, the number of connected
components $\beta_0$ for each of these $\epsilon$ values become the
elements of the column vector for that time point in the CROCKER
matrix.  An example CROCKER plot for the first 900 seconds of one of
the experimental trials is shown in the top panel of
Fig.~\ref{fig:exp_crocker}.
As before, this plot gives clear visual evidence of the phase
change in the structure that follows the introduction of the donor
bees.

To identify this phase change using our methodology, we begin, as
before, by reducing each column vector $\vec{b}(t)$ of the CROCKER
matrix down to a scalar value using the $\ell^2$ norm, then apply the
{\tt changepoint}, $k$-means, and agglomerative methods to the
resulting $||\vec{b}(t)||$ time series.  Since the goal, again, is to
identify a single phase change, we set the algorithms up to search
for one change-point or two clusters.  The change-point algorithm
flagged $t_{shift}=475$ in this experiment: {\sl i.e.}, 45 seconds
after the introduction of the donor bees at $t=430$.
The two clustering algorithms did not produce clean results, however.
Unlike in the simulation experiments, they did not find clearly
delineated clusters.  Rather, the overlap region, where successive
points are classified in different phases, spanned more than 90\% of
data set: an average width of 845.0 and 830.2 seconds, respectively,
for $k$-means and agglomerative clustering.

The inability of the clustering algorithms to clearly distinguish the
morphological phases in these data---which are quite visible to the
eye, both in the videos and in the CROCKER plots---is not surprising,
as those plots are far noisier than the ones constructed from the
simulation data.  In the real world, we often find stray bees
(sometimes dead bees) who do not participate in the clustering.  As a
result, the contrast between the clustering and non-clustering phases
in the CROCKER plot is not very drastic, as observed by the contours
in the two phases.  In the simulation data, we do not see this effect;
rather, the bee-agents move quickly to form clusters and thus the
phase change is clear and drastic.  To use these clustering algorithms
to identify the phase changes in the face of these difficulties, we
have to employ a different strategy.  Recall that working only on the
norms effectively discards the temporal information about each of the
values passed to the clustering algorithms.  (This is not at issue for
the R {\tt changepoint} algorithm, which treats its input as a time
series.)  To regain the fact that the data is a time series, we can
annotate each norm value with the time step to which it corresponds,
yielding tuples of the form $[\, ||\vec{b}(t_i) ||, t_i]$ where $t_i$
is the time stamp and $||\vec{b}(t_i) ||$ is the $\ell^2$ norm of the
vector of $\beta_0$ values of the associated filtration.  This greatly
improves the results, reducing the mean widths of the overlap region
to 1.667 and 12.167 for $k$-means and agglomerative clustering,
respectively.
For the CROCKER matrix example in Fig.~\ref{fig:exp_crocker}, the
agglomerative algorithm, working with the time-augmented data, flagged
the phase change 17 time steps after the introduction of the donor bees
at $t=430$ ({\sl i.e.}, $t_{shift}=447$) while the $k$-means algorithm
  signalled that change at $t=465$.

An analysis of the $t_{shift}$ results across multiple trials is
somewhat more complicated here than in the model runs of
Section~\ref{sec:tda} because the introduction time of the donor bees
is different for each laboratory experiment.  To account for this
variation, we calculate the number of time steps $t_{lag}$ between the
introduction of the donor bees ($t_{donor}$) and the change detected by
the clustering algorithms ($t_{shift}$) for each experiment and then
compute the summary statistics on $t_{lag}=t_{shift}-t_{donor}$.
Across six runs of the experiment, the means and standard deviations
of $t_{lag}$ were $[58.33, 115.0]$ and $[87.2,72.0]$, respectively,
for the $k$-means and agglomerative clustering algorithms.
A primary cause of the large standard deviation in both cases was a
single experimental trial (C0133) in which the bees formed a
short-lived aggregation before the introduction of the donor bees.
The implications of this are discussed at more length in
Section~\ref{sec:discussion}.  Across all six experimental runs, the
mean and standard deviation of the $t_{lag}$ values produced by the
change-point detection algorithm were 71.3 and 15.5.  Across all three
algorithms and all six experiments, the $t_{lag}$ values were later
for the experimental data than the model runs, suggesting that the
delay between introduction of the donor bees and the formation of
aggregations is longer in the experiment than in the simulation.  We
discuss possible reasons for these observations in the following
section.

\section{Discussion}
\label{sec:discussion}

In both simulated and laboratory data, the rich morphological
signature produced by persistent homology provided leverage to
change-point detection and clustering algorithms for identifying phase
changes in the behavior of honeybee aggregations.  Bees exemplify a
significant application of this TDA-based approach, given the vital
role that aggregations play in their biology and behavior.  The
temporal evolution of the topological signature, as captured in
CROCKER matrices, is potentially relevant to the study of other
aggregation-related behaviors in bees, including social-network
analysis in the context of disease \cite{geffre2020honey}, waggle
dance communication \cite{seeley2011honeybee, dong2023social}, and the
aggregation around the queen through chemical communication
\cite{nguyen2021flow}.  The utility of this methodology encompasses
other social insects, such as ants and their trophallaxis behavior
\cite{baltiansky2023emergent}.  Given its generality, the
applicability of our TDA-based methodology extends beyond biological
aggregations: it could be used to identify phase changes in any other
dynamically evolving point cloud.

\begin{figure}
\begin{center}
\includegraphics[width=0.8\linewidth]{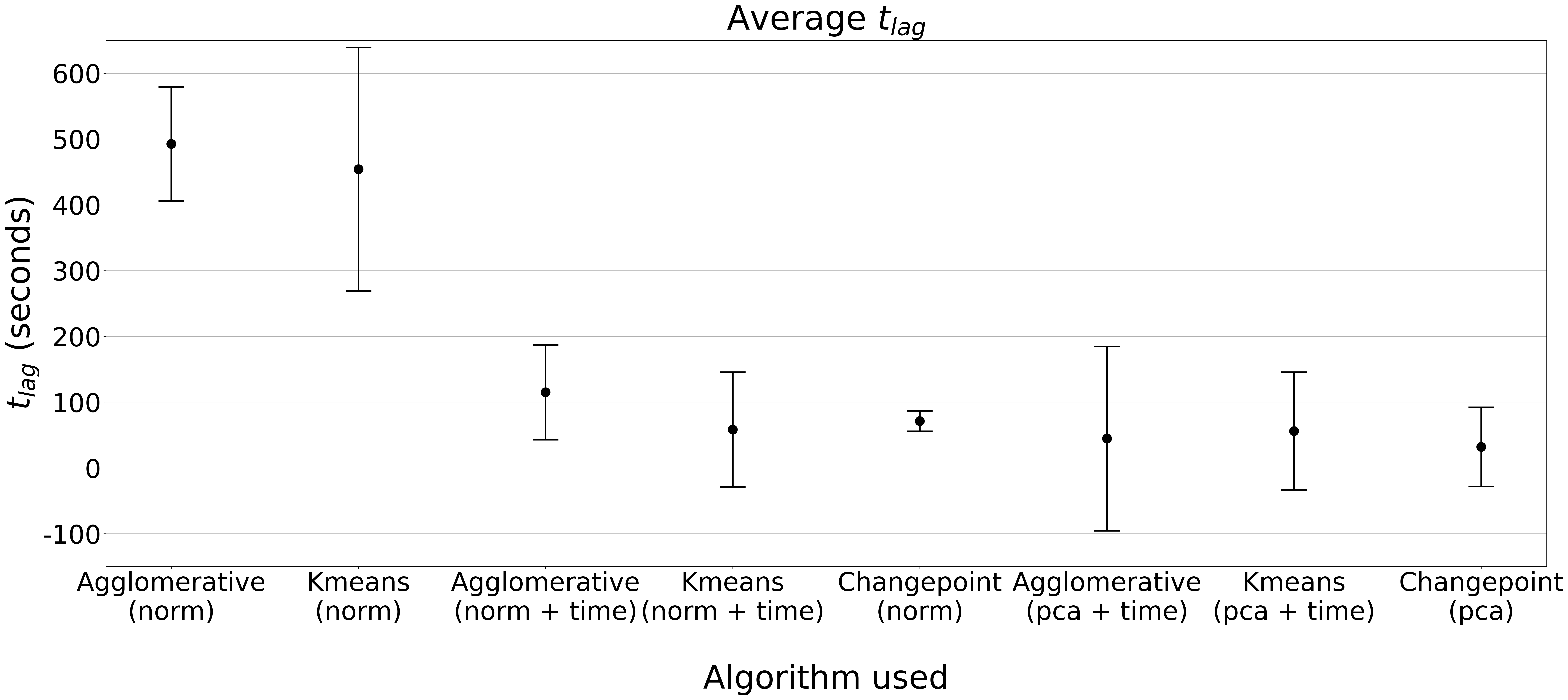}
\end{center}
\caption{Survey of results of all phase-shift detection approaches
  across all six experimental data sets, showing means (blue bars) and
  standard deviations (black whiskers) for $t_{lag}$, measured in
  seconds after the introduction of the donor bees.  Note that the
  {\tt changepoint} algorithm inherently considers the input as a time
  series, whereas the clustering algorithms do not. }
\label{fig:all}
\end{figure}

The results of all combinations of the different algorithms and
dimensional-reduction techniques described in the previous sections
are depicted graphically in Fig.~\ref{fig:all}.
(Please see the Supplementary Material for a table of the $t_{lag}$
values for the individual experiments.)  The four leftmost bars bring
out the previously noted inability of the two clustering algorithms to
identify the phase change in the absence of any information about
time.  This is not surprising because of the noise in the data, which
makes the change in the contours of the CROCKER plot far less abrupt.
Explicitly re-introducing time as part of the input to these
algorithms made a clear difference in the results.  Notably, all of
these experimental $t_{lag}$ values are larger than in the
simulations.  That is, the bees appeared to form aggregations much
more quickly in the model than in the laboratory: $22.8 \pm 12.9$,
$29.0 \pm 27.2$, and $22.2 \pm 12.9$ for $k$-means, agglomerative, and
{\tt changepoint}, respectively---compared to $58.3 \pm 87.2$, $115.0
\pm 72.0$, and $71.3 \pm 15.5$ for the experimental data (augmented
with time, in the case of the two clustering algorithms).  There are
two likely reasons for this.  First, the model time scales are not
identical to those of the experiments.  A calculation of the average
speed in the two settings suggests a factor of five difference in
those time scales, with the model being faster---which is consistent
with the lower $t_{lag}$ values.
Second, the model rules are only an approximation of what the bees
really do.  Importantly, those rules do not include scenting
behaviors, which bees use to communicate information about, for
example, the presence of food.  This, too, could affect the time
scales of the aggregation behavior.

It is perhaps surprising that reducing the dimension of the
multi-scale topological signature down to a scalar time series using a
method like a norm---which distills each $n$-element CROCKER vector
down into a single number---leaves enough information for algorithms
to detect the phase change, but others have observed similar effects
\cite{munch2022}.  There are other dimension-reduction techniques, of
course: notably principal component analysis or PCA.  To explore this
alternative, we repeated the clustering analysis by projecting each
$\vec{b}(t_i)$ onto the first principal component of the overall
CROCKER matrix in Fig.~\ref{fig:exp_crocker}, annotated with the
associated time stamp: {\sl i.e.}, using the $k$-means and
agglomerative algorithms to seek two clusters in a series of
two-vectors of the form $[c_{\vec{b}(t_i)},t_i]$, where
$c_{\vec{b}(t_i)}$ is the projection of $\vec{b}(t_i)$ onto the first
principal component of the entire matrix.  The results, respectively,
were $t_{lag}=56.2 \pm 89.5$ and $t_{lag}=44.7 \pm 140.0$.
In both cases, the standard deviation was skewed by the experimental
trial mentioned above (C0133) in which the bees formed a small,
short-lived cluster before the introduction of the donor bees; the
agglomerative case was additionally skewed by an early $t_{shift}$
detection in a second data set (C0128), where a similar but smaller
and shorter-lived aggregation formed before $t_{donor}$.  Repeating
the analysis on the $c_{\vec{b}(t_i)}$ values with the {\tt
  changepoint} method, we obtained $t_{lag}=10.5 \pm 120.1$ across the
six experimental trials.  Here, too, the C0133 data set was the source
of the large $\sigma$.  A lateral comparison of the norm- and
PCA-based results across all three algorithms and all six experiments
suggests that the former produces earlier mean $t_{lag}$ values, but
with much larger variability across experiments.  (In the C0133 trial,
for instance, {\tt changepoint} flagged a negative $t_{lag}$ when
working with the first principal component and a positive one when
given the norm trace.)  From both practical and theoretical
standpoints, we prefer the norm-based method because it is both
simpler and also more temporally precise.  The principal components
computed by PCA incorporate information from the entire CROCKER
matrix---explaining the variance of the $\vec{b}(t)$ vectors for every
time point in the experiment---whereas $||\vec{b}(t_i)||$ is specific
to a given time point.

Assessing these results, again, is made difficult by the fact that we
do not have ground truth for the phase change.  {\sl
  Trophallaxis-induced} aggregations should certainly not form before
the introduction of the donor bees, but aggregations can form for
other reasons in groups of bees: because of generalized attraction
between individuals, chemical information exchange
\cite{leboeuf2016oral}, or even just spatial inhomogeneities that
emerge naturally during random walks \cite{spitzer2013principles}.
Close examination of the experimental videos, available in the
Supplementary Materials, shows small, transient aggregations forming
and dispersing before $t_{donor}$ in all six trials.  In the two that
caused the different algorithms to flag a negative $t_{lag}$ (C0133
and, to a lesser extent, C0128) these aggregations were simply
somewhat larger and longer-lived.  The aggregations that formed {\sl
  after} the introduction of the donor bees were far larger and
longer-lived, as is clear from the contours in the CROCKER plots and
the means in Fig.~\ref{fig:all}.  In other words, the TDA-based
techniques effectively bring out both the large-scale phase
changes and some of the nuances of the behavior.

To study the salience of the different elements of the topological
signature for the purposes of clustering the data into these distinct
behavioral phases, we use a Chi-square statistical test.  To each data
point, we assign the label produced by the two clustering algorithms.
Then we treat the full $\vec{b}$ vector for each time step as a
feature vector and determine which of its elements---{\sl i.e.}, the
$\beta_0$ values for individual $\epsilon$ values---is the most highly
correlated with those labels.  Performing a Chi-square test in this
way allows us to reverse-engineer the ``most influential'' values of
the $\epsilon$ parameter for each trial.  Across all six experiments,
the mean of these values was $117.3 \pm 27.9$ pixels (1.3 $\pm$ 0.3 in
units of bee body lengths).  While it would be inappropriate to impute
physical meaning to this result ({\sl e.g.}, mechanisms of honeybee
behavior), it does provide a major potential advantage, since it means
that one need not build the full series of filtrations, but rather
just a few in that range.

A related matter here is the notion of approximating bees as points.
Since this effectively neglects their actual spatial extent, it
introduces a systematic downward bias in the number of
$\epsilon$-connected components for a given $\epsilon$ value, since
the bee perimeters, which are what really define whether the animals
are close, will generally be closer than their centers.  A proper
study of the effects of this would be a real challenge, requiring the
development of novel computer-vision methods (to extract the actual
perimeters) and novel TDA techniques that move beyond the fundamental
assumption of {\sl point}-cloud data, perhaps using oriented
ellipsoids to define connectivity.  However, the results in the
previous paragraph---that the ``most influential'' $\epsilon$ value is
larger than the length of a bee---suggest that approximating bees as
points is not a major issue in our approach.  

The analysis scales in the construction of the topological signature
are an interesting matter here.  Fundamentally, TDA is not about the
points, but rather about the spaces between them.
In experimental data, the natural metric for those spaces---and thus
the appropriate unit for the TDA calculations---is dictated by the
measurement apparatus: in our laboratory experiments, the pixel
resolution of the video camera.  These may, of course, differ in other
experimental setups, but there are some standard procedures for
setting up the filtration that make this kind of analysis systematic.
As mentioned in Section~\ref{sec:tda}, it is common to set the lower
bound $\epsilon_{min}$ of the range of the scale parameter so that
every point is a single component\footnote{though obviously not below
the resolution of the data}.  Resolution limits can, of course, create
spurious topological effects: if the pixels recorded by the camera are
5mm on a side, for instance, two points that are separated by 1mm will
be treated as touching even though they are not.  (This is not a
shortcoming of TDA, of course, but rather a general issue with data
resolution.)  The value of $\epsilon_{max}$ is also generally dictated
by the data.  Since increasing $\epsilon$ beyond the value that
connects all of the points into a single $\epsilon$-connected
component will not add to the information in the topological
signature, it makes sense to take $\epsilon_{max}$ such that the
entire set is connected, as we do in both our real and synthetic data
sets.  Between those limits, the number of steps in the filtration
dictates the resolution of the topological signature: {\sl i.e.,} how
precisely one knows what $\epsilon$ value connects the points into a
particular number of $\epsilon$-connected components.  Since the
computational cost of TDA rises with the number of complexes in the
filtration, this choice can involve balancing a tradeoff between
computational complexity and analysis resolution.  This cost is modest
in our data sets, which contain tens of points, so we choose the
smallest spacing $\Delta \epsilon$ of the filtration parameter that is
available in the experimental data: one pixel, which is roughly 1/55th
of the average body length of the bees in these images.  Different
data sets, with different numbers of points and/or different spatial
resolutions, may require different choices for $\Delta \epsilon$ and
for $[\epsilon_{min}, \, \epsilon_{max}]$.  While that will rescale
the vertical axis of the associated CROCKER plot, it will not obscure
the information that it contains, nor will it affect the performance
of our proposed methods.

\begin{figure}
\begin{center}
\includegraphics[width=0.6\linewidth]{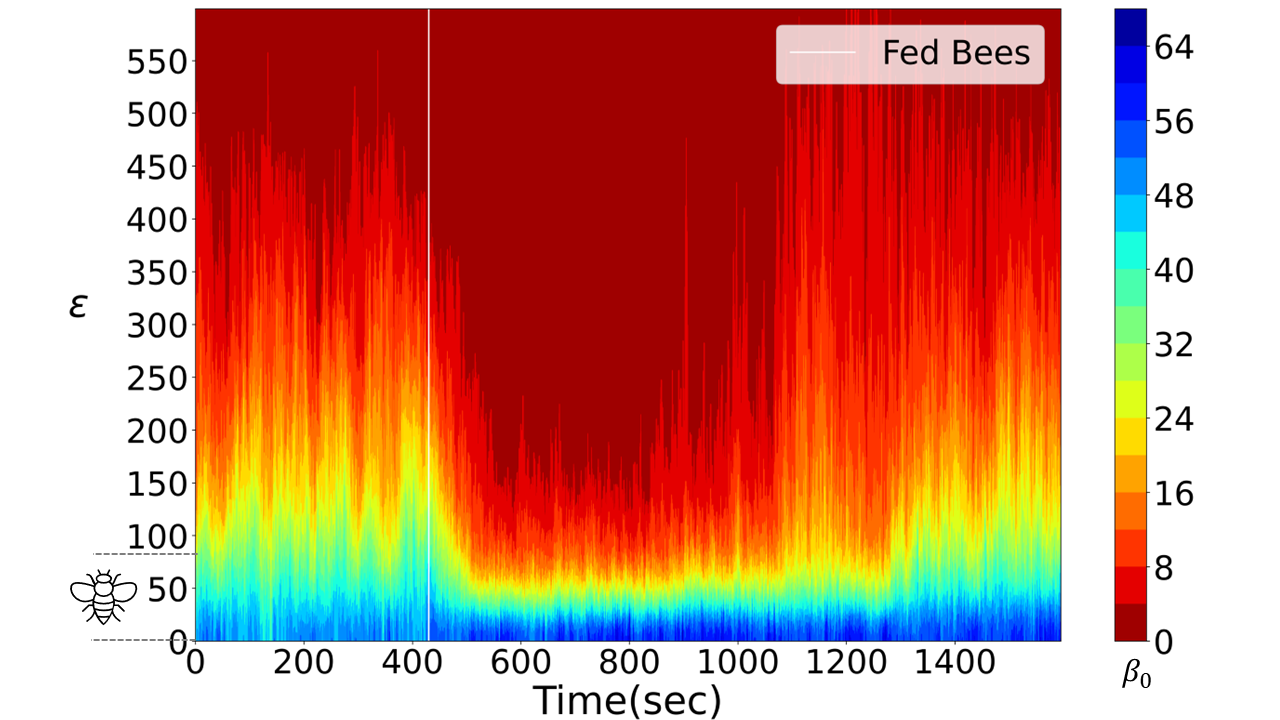}
\\
\medskip
\includegraphics[width=0.6\linewidth]{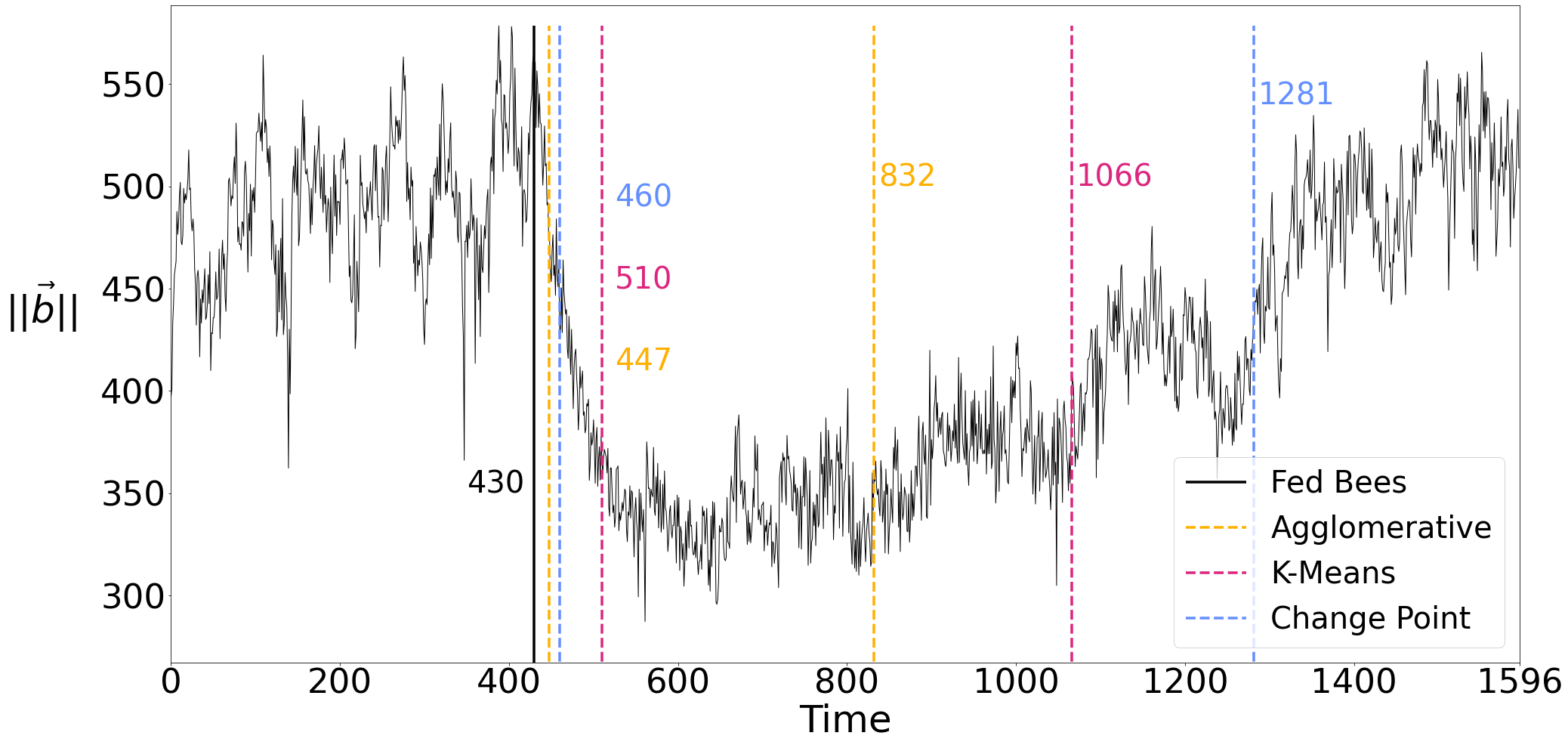}  
\end{center}
\caption{A CROCKER plot of a longer experiment (top) reveals a second
  phase change in the morphology, likely at the point where the food
  is distributed evenly across the group.  Results of the {\tt
    changepoint} (blue), $k$-means (red), and agglomerative (orange)
  algorithms applied to the $[\, || \vec{b}(t_i) ||, t_i]$ vectors,
  shown superimposed on the bottom image, corroborate this result.  As
  before, the black solid line in the bottom plot marks the time at
  which the donor bees are introduced.}
\label{fig:longer-run}
\end{figure}

If food exchanges play a role in the formation of aggregations in a
group of honeybees, it is reasonable to explore what happens in longer
experiments, when the food has diffused across the group and exchanges
are presumably less frequent.  Fig.~\ref{fig:longer-run} shows a
CROCKER plot for such an experiment, along with the associated norms
and clustering results.
Interestingly, this experiment reveals an additional change in the
morphology towards the end of the experiments---though the three
different algorithms flagged that change at quite different times.
(Note that the $t_{lag}$ locations for the first phase change detected
by the two clustering algorithms are different than in
Fig.~\ref{fig:exp_crocker} because the data set in this experiment
contains many more CROCKER vectors, which shifts the geometry of the
clusters.)  We hypothesize that this third dispersed phase, whose
morphology resembles the first one, occurs when the food is
distributed evenly across the group and trophallaxis events play less
of a role in the behavior of the bees, causing them to gradually break
the aggregations and return to a dispersed, random motion pattern,
similar to the first phase.  However, additional experiments are
required to confirm these observations and hypotheses.

Lastly, our approach has the potential to bring a deeper understanding
to collective animal behavior cell aggregations, to insect swarms,
bird flocks, and fish schools.  Our methodology could be easily
extended beyond aggregations as well: e.g., by constructing CROCKER
plots from $\beta_1$, which counts the number of holes in a point
cloud, we could study milling behaviors in honeybee swarms and sheep
herds.  It can also be applied to higher-dimensional data, such as
three-dimensional schools of fish and flocks of birds.  Additionally,
there is room for enhancement in the methodology itself, such as
refining the computer-vision code used to detect the positions of the
bees, exploring alternative clustering and change-point detection
algorithms, and incorporating other dimension-reduction techniques.
Another important potential affordance of this methodology is model
validation via a comparison of topological signatures of simulation
and experiment \cite{ulmer2019topological}.

\section{Conclusion}
\label{sec:conclusion}

In this paper, we have proposed a new TDA-based method for detecting
phase changes in biological aggregations and demonstrated it on data
from honeybees, first in the context of an agent-based model and then
using data from laboratory experiments.  Persistent homology has been
used extensively in the past decade for analyzing the structure of
many different kinds of data, ranging from point clouds to images, but
identifying phase changes in point-cloud structure has received
less attention.  CROCKER plots are an effective visual representation
of how topological signatures change over time.  To leverage that
information for the purposes of identifying phase changes, we
compress the information in the CROCKER plot into a scalar time
series.  We used two different dimension-reduction strategies for
this: (i) simply taking the $\ell^2$ norm of the CROCKER vectors
$\vec{b}(t)$ at each time point (ii) performing PCA on the whole
CROCKER matrix and then projecting each $\vec{b}(t)$ onto the first
principal component.  We tested three algorithms on the resulting
data: a traditional change-point detection algorithm and two standard
clustering algorithms.  We demonstrated these approaches on simulated
and experimental data sets of honeybee movement, with the goal of
detecting the phase change that follows the introduction of donors
into a group of hungry bees, as the individuals exchange food and
aggregations form.

Our approach is novel from the perspective of detecting phase changes
in evolving point clouds, and the alignment of the results from these
approaches indicates their success.  All of the different strategies
for algorithm and dimensional reduction produce roughly similar
results, detecting the phase change within 50-100 seconds from the
time when the donor bees are introduced.  Bees are by no means the
only application for these strategies; our method can be used to track
phase changes in any point cloud, regardless of its provenance, and
aid in the understanding of scientific processes that are involved in
the dynamics of those data, as well as in validating models of those
processes.  Overall, we hope that this methodology will help advance
our understanding of biological aggregations and their intricate
dynamics.

\section{Methods}
\label{sec:methods}

Methods necessary for the replication of the results are comprehensively
described in Section~\ref{sec:tda} and Section~\ref{sec:experiments}. 

\section{Data availability}
\label{sec:data_availability}

The raw image files for dataset C0128 analysed during the current study 
is available in the GitHub repository, 
\url{https://github.com/vrd1243/tda_bees_datasets}. The raw image files
for the remaining datasets analysed during the current study are not 
publicly available due to data limitations, but are available from the 
corresponding author on reasonable request. Sped-up versions of the 
video files for the six experimental datasets, and the video
and the text files of the trophallaxis simulation run used in the 
paper are available at \url{
  https://github.com/vrd1243/tda_bees/} under the {\tt data/} subdirectory.

\section{Code availability}
\label{sec:code_availability}

The underlying code for this study is available in \url{
  https://github.com/vrd1243/tda_bees/}.

\bibliographystyle{unsrt}
\bibliography{refs}

\section{Acknowledgments}

The authors gratefully acknowledge useful conversations with James
Meiss and Joshua Garland, as well as funding from the National Science
Foundation grants CMMI 1537460 and AGS 2001670, and the BioFrontiers Institute at the University of
Colorado Boulder.  Any opinions, findings, and conclusions or
recommendations expressed in this material are those of the authors
and do not necessarily reflect the views of the National Science
Foundation.

\section{Author Contributions}

G.G.F. performed experiments; G.G.F., M.B., V.D., C.M., and C.T. performed agent-based modeling and topological data analysis; C.M. performed the change-point analysis; M.B., and V.D. performed the dimensionality reduction and clustering analysis; G.G.F., V.D., and E.B. wrote the paper with inputs from M.B., O.P., C.M., and C.T.; E.B., C.T., and O.P. designed and supervised research.

\section{Competing Interests}

All authors declare no financial or non-financial competing interests. 

\end{document}



{\noindent \large \bf Supplementary Material for ``A Computational Topology-based Spatiotemporal Analysis
  Technique for Honeybee Aggregation''}

\bigskip
Golnar Gharooni-Fard*, Morgan Byers*, Varad Deshmukh*, Elizabeth
Bradley, Carissa Mayo, Chad Topaz, Orit Peleg \\ * contributed equally
\bigskip

\section{Summary table of $t_{lag}$ values in Fig. 8}

We presented a comparison of the results from all phase-shift detection approaches across
all the six datasets in the form of dot-whisker plots in Fig. 8. Since only six data points 
were available for each approach, we summarize the individual $t_{lag}$ values in 
Tbl.~\ref{tbl:tlag}.
\bigskip

\begin{table}[htbp]
\begin{tabular}{|c||c|c|c|c|c|c|}\hline
    & \multicolumn{6}{c|}{\textbf{Dataset $t_{lag}$ value in seconds}} \\
    \hline
    \textbf{Algorithm} & \textbf{C074} & \textbf{C077} & \textbf{C081} & \textbf{C0128} & \textbf{C0132} & \textbf{C0133} \\
    \hline
    \hline
    \textbf{Agglomerative} (norm) & 638 & 480 & 492 & 469 & 509 & 368 \\
    \textbf{Kmeans} (norm) & 638 & 479 & 542 & 469 & 499 & 98 \\
    \textbf{Agglomerative} (norm + time) & 115 & 92 & 186 & 17 & 209 & 71 \\
    \textbf{Kmeans} (norm + time) & 184 & 32 & 105 & 35 & 72 & -78 \\
    \textbf{Changepoint} (norm) & 69 & 66 & 89 & 45 & 76 & 83 \\
    \textbf{Agglomerative} (pca + time) & 191 & 186 & 103 & -159 & 2 & -55 \\
    \textbf{Kmeans} (pca + time) & 188 & 37 & 102 & 21 & 70 & -81 \\
    \textbf{Changepoint} (pca) & 113 & 69 & 10 & 25 & 69 & -223 \\
    \hline
\end{tabular}
\caption{Summary table of the values for the various phase-shift detection
methods across the six datasets.} 
\label{tbl:tlag}
\end{table}